\documentclass[sigconf]{acmart}

\usepackage[english]{babel}
\usepackage{blindtext}
\usepackage[ruled]{algorithm2e}
\usepackage{changepage}
\usepackage{enumitem}
\usepackage{pdfpages}
\usepackage{subfigure}
\usepackage{tabularx}
\usepackage{multirow}
\usepackage{array}
\usepackage{diagbox}
\usepackage{graphics}
\usepackage{hyperref}
\usepackage{wasysym}
\usepackage{comment}
\hypersetup{
    colorlinks=true,
    linkcolor=blue,
    filecolor=magenta,      
    urlcolor=cyan,
    pdftitle={Overleaf Example},
    pdfpagemode=FullScreen,
    }

\AtBeginDocument{%
  \providecommand\BibTeX{{%
    \normalfont B\kern-0.5em{\scshape i\kern-0.25em b}\kern-0.8em\TeX}}}

\copyrightyear{2023}
\acmYear{2023}
\setcopyright{acmlicensed}
\acmConference[MM '23]{the 31st ACM International Conference on Multimedia}{October 29-November 3, 2023}{Ottawa, ON, Canada}
\acmPrice{15.00}
\acmDOI{10.1145/3581783.3612177}
\acmISBN{979-8-4007-0108-5/23/10}

\begin{document}
\begin{sloppypar}
\title{Karma: Adaptive Video Streaming via Causal Sequence Modeling}

\author{Bowei Xu}
\affiliation{%
  \institution{Nanjing University}
   \city{Nanjing}
   \country{China}
}
\email{xubowei@smail.nju.edu.cn}

\author{Hao Chen}
\affiliation{%
  \institution{Nanjing University}
   \city{Nanjing}
   \country{China}
}
\email{chenhao1210@nju.edu.cn}

\author{Zhan Ma}
\affiliation{%
  \institution{Nanjing University}
   \city{Nanjing} 
   \country{China}
}
\email{mazhan@nju.edu.cn}


\begin{abstract} 

Optimal adaptive bitrate (ABR) decision depends on a comprehensive characterization of state transitions that involve interrelated modalities over time including environmental observations, returns, and actions. However, state-of-the-art learning-based ABR algorithms solely rely on past observations to decide the next action. This paradigm tends to cause a chain of deviations from optimal action when encountering unfamiliar observations, which consequently undermines the model generalization.

This paper presents Karma, an ABR algorithm that utilizes causal sequence modeling to improve generalization by comprehending the interrelated
causality among past observations, returns, and actions and timely refining action when deviation occurs. Unlike direct observation-to-action mapping, Karma recurrently maintains a multi-dimensional time series of observations, returns, and actions as input and employs causal sequence modeling via a decision transformer to determine the next action. In the input sequence, Karma uses the maximum cumulative future quality of experience (QoE) (a.k.a, \textit{QoE-to-go}) as an extended return signal, which is periodically estimated based on current network conditions and playback status. We evaluate Karma through trace-driven simulations and real-world field tests, demonstrating superior performance compared to existing state-of-the-art ABR algorithms, with an average QoE improvement ranging from 10.8$\%$ to 18.7$\%$ across diverse network conditions. Furthermore, Karma exhibits strong generalization capabilities, showing leading performance under unseen networks in both simulations and real-world tests. 

\end{abstract}



\maketitle

\section{Introduction}
\label{sec:intro}

In recent years, a remarkable surge of HTTP-based video traffic~\cite{index2015cisco,cisco2017cisco} has been propelled by the proliferation of video streaming applications. Adaptive Bitrate (ABR) algorithms~\cite{dobrian2011understanding, krishnan2012video,bentaleb2018survey} have emerged as prominent tools and have been employed by content providers to optimize the video quality of streaming services. Typically implemented on the client-side video player, the ABR algorithm dynamically adjusts the video bitrate in response to the underlying network conditions, with the primary objective of maximizing users' quality of experience (QoE).  

\begin{figure}[t]
    \centering
    \setlength{\abovecaptionskip}{0cm} 
    \subfigure[Existing schemes via observation-to-action mapping]
    {\includegraphics[scale = 0.85]{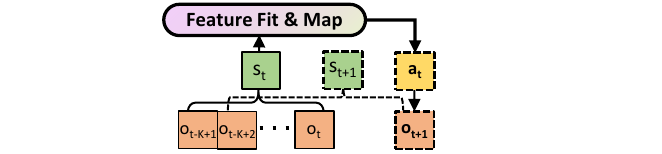}
    \label{sfig:problem_fomulation_existingABR}}\\
    \vspace{-0.2cm}
    \subfigure[Karma via causal sequence modeling]{\includegraphics[scale = 0.85]{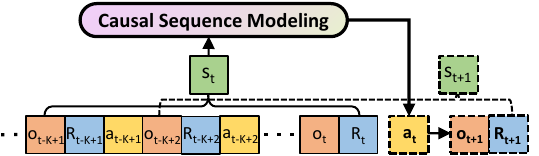}
    \label{sfig:problem_fomulation_Karma}}
    \caption{{\it ABR Decision In Use} : Karma vs. existing algorithms.}
    \label{fig:problem_fomulation}
    \vspace{-0.2cm}
\end{figure}

Early rule-based ABR algorithms rely on fixed control rules for bitrate decisions. However, they usually require careful tuning (e.g., rate-based~\cite{sun2016cs2p,jiang2012improving} and buffer-based algorithms~\cite{huang2014buffer,spiteri2020bola}) or prior knowledge of network conditions (e.g., MPC~\cite{yin2015control}), making them incapable of being generalized well in most dynamic networks~\cite{huang2012confused,sun2016cs2p,zou2015can}. Although learning-based ABR approaches (e.g., imitation learning (IL)~\cite{osa2018algorithmic,huang2019comyco} and reinforcement learning (RL)~\cite{jaderberg2016reinforcement,sutton2018reinforcement,mao2016resource,mao2017neural} algorithms) have shown superior performances, they
solely stack past observations to decide the next action, as illustrated in Figure~\ref{sfig:problem_fomulation_existingABR}. When familiar observations cannot be encountered in unexperienced environments, these learning-based algorithms tend to fall into a chain of deviations from optimal actions, which ultimately undermines the model generalization~\cite{de2019causal, xia2022genet}. A detailed analysis of existing ABR algorithms and their limitations are revealed in §\ref{sec:background}.

This paper proposes Karma, a novel ABR decision system that aims to enhance generalization by comprehending the interrelated causality among past observations, returns, and actions and executing timely action refinement when deviation occurs. As shown in  Figure~\ref{sfig:problem_fomulation_Karma}, it recurrently uses a sequence of multi-dimensional elements, including observations, returns, and actions over time as the state input to determine the next action. In Karma, The return signal is formulated as a ``QoE-to-go'' representing the {\it maximum cumulative future QoE} to better assess the current state from a long-term perspective. To thoroughly characterize the cross-time and inter-modality causality in each input sequence, i.e., causal sequence modeling, for better action decisions (e.g., bitrate of the next video chunk), Karma proposes the use of causal decision transformer~\cite{vaswani2017attention,chen2021decision} to fulfill the purpose. We describe the principle and superiority of the proposed causal sequence modeling in §\ref{sec:sequential_modeling}. 

\begin{figure}[t]
    \centering
    \setlength{\abovecaptionskip}{0.1cm} 
    \subfigure[IL-based ABR algorithms]
    {\includegraphics[scale = 0.22]{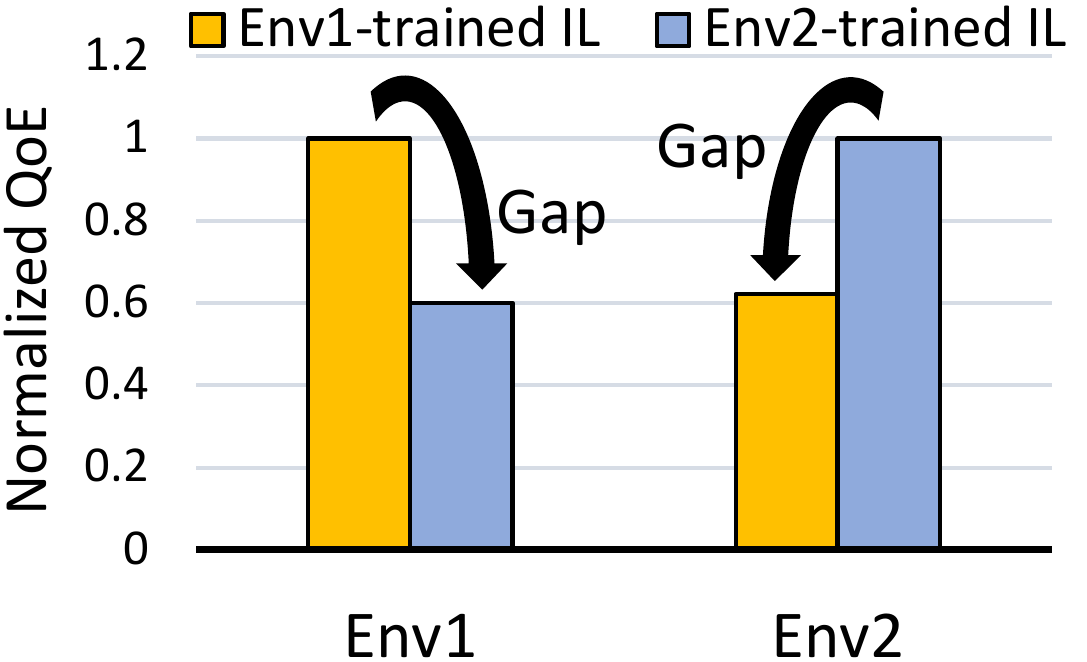}
    \label{sfig:IL_generalization_problem}}
    \subfigure[RL-based ABR algorithms]{\includegraphics[scale = 0.22]{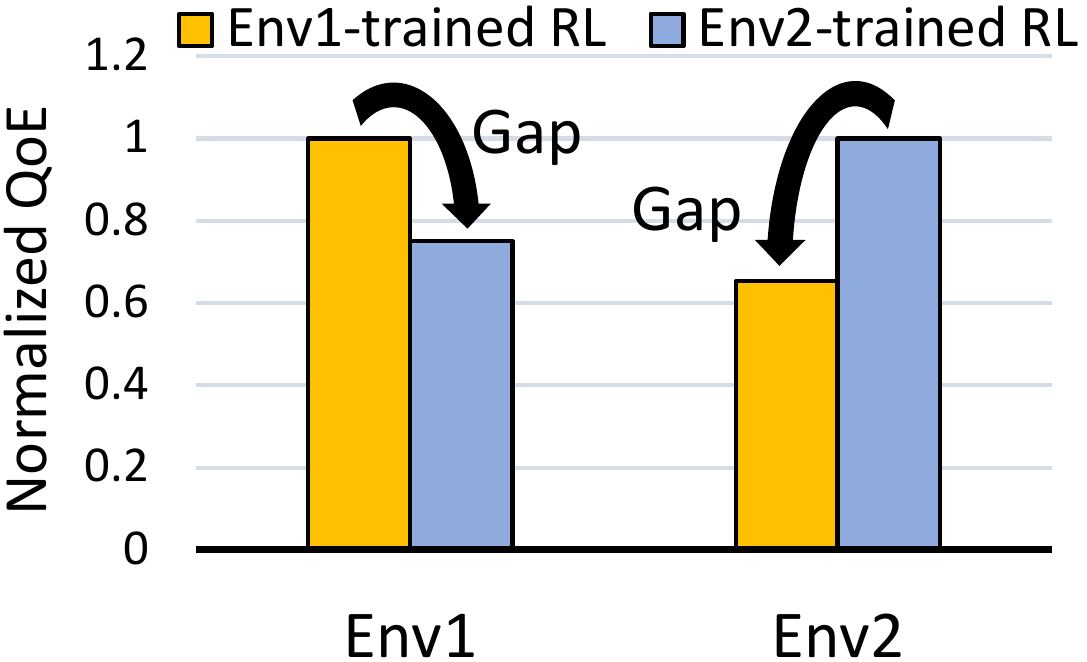}
    \label{sfig:RL_generalization_problem}}
    \vspace{-0.1cm}
    \caption{Generalization issues of both IL- and RL-based ABR algorithms. Env1 and Env2 have an average bandwidth of 1.3Mbps and 2.0Mbps, respectively.  Results are normalized using the model performance measured in the same training environment.}
    \label{fig:observation_generalization}
    \vspace{-0.5cm}
\end{figure}

To train Karma, we first construct training samples by generating numerous extended expert ABR trajectories. Each trajectory comprises serial chunk-level tuples, each of which consists of observation, estimated QoE-to-go, and optimal action at each chunk. A pair of observations and optimal action for each chunk is obtained via dynamic programming. It serves as the expert guide for the pursuit of optimality. And the corresponding QoE-to-go is estimated using the current network condition and playback status via a pre-trained estimator. Using these expert ABR trajectories, Karma trains a causal decision transformer via supervised learning by minimizing the cross-entropy between the optimal actions (labeled in trajectories) and Karma predictions. We describe the design and implementations of Karma in §\ref{sec:design}.

Karma is compared with state-of-the-art ABR algorithms under a wide range of network conditions in both trace-driven simulations and real-world field tests.  Our results indicate that Karma achieves an average QoE improvement ranging from 10.8$\%$-18.7$\%$ over the best-performing solution in each scenario under consideration (§\ref{ssec:overall_performance}). It is also worth mentioning that Karma consistently outperforms prevalent learning-based ABR algorithms under networks that have never been experienced in both simulations and real-world tests (§\ref{ssec:generalization_peformance}). All of these studies not only report the superior performance of Karma but also reveal its wide generalization. Additional deep dive studies are also conducted to offer in-depth insights into Karma (§\ref{ssec:ablation_peformance}).

In general, we summarize our contributions as follows:
\vspace{-0.1cm}

\begin{itemize}[leftmargin=*]
    \item To the best of our knowledge, Karma is the {\it very first solution} that recurrently maintains a multi-dimensional time series of observations, returns, and actions and employs causal sequence modeling via a decision transformer to determine the next action in the ABR system, which significantly ameliorates the weakness in generalization for existing learning-based approaches that solely rely on the observation-to-action mapping;
    \item We utilize a maximum cumulative future QoE (i.e., QoE-to-go) as an effective return signal to reflect the long-term assessment of the current state and devise a learning-based QoE-to-go estimator in Karma, which is also different from the instant returns used in existing approaches;  
    \item Extensive experimental results in both trace-driven simulations and real-world field tests demonstrate the proposed Karma's remarkable generalization and superior performance across a broad set of network conditions.	 	
\end{itemize}

\vspace{-0.1cm}
\section{Background and Motivation}
\label{sec:background}
HTTP-based adaptive streaming is a dominant tool used to deliver video content across the entire Internet today. Relevant techniques have also been approved as the international standard like DASH~\cite{dashjs} to assure service interoperability across various heterogeneous users. As the underlying network conditions usually fluctuate unexpectedly, ABR algorithms are usually devised to sustain uncompromised QoE.

\begin{figure}[t]
\centering
\setlength{\abovecaptionskip}{0.1cm} 
\includegraphics[scale = 0.24]{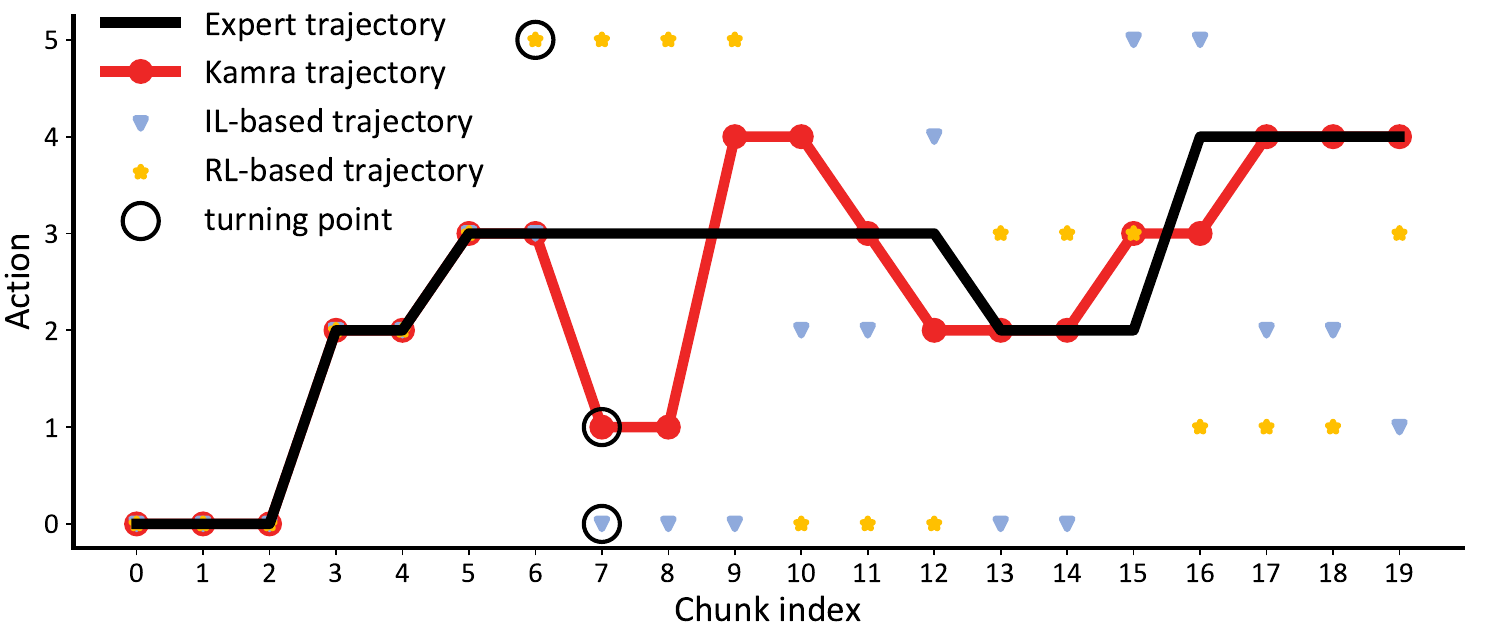}
\caption{Comparing the action trajectories using the expert policy, the IL-based model, the RL-based model, and our proposed Karma.}
\label{fig:trajectory_comparison}
\vspace{-0.5cm}
\end{figure}

Existing ABR algorithms can be grouped into two classes: rule-based algorithms and learning-based algorithms. As extensively analyzed in~\cite{mao2017neural}, rule-based algorithms cannot easily generalize themselves to different networks because deliberate tuning of parameters or a precise understanding of system dynamics is required. 
Unfortunately, unsatisfactory generalization is also troubling recent learning-based ABR algorithms. Xia et al.~\cite{xia2022genet} reported that existing learning-based ABR algorithms tend to better adapt to the same environment used in training, and significant degradation is observed when they operate in a new environment. The same observation is also reproduced in our preliminary experiment.
We prepare two separate environments (i.e., Env1 and Env2) by dividing a part of network traces from Norway 3G/HSDPA mobile dataset~\cite{riiser2013commute} into two corpora with different average bandwidths. 
Representative IL- and RL-based ABR models are respectively trained in Env1 and Env2 and then tested in both of them as well.
As shown in Figure~\ref{fig:observation_generalization}, significant performance degradation is observed when testing the model in a different environment for both IL- and RL-based approaches. More details will be provided in §\ref{sec:evaluation}.

To investigate the reason for this generalization issue, we further analyze the action trajectories generated using the expert policy, the IL-based ABR model (Comyco), the RL-based ABR model (Pensieve), and the causal sequence modeling based ABR model (our proposed Karma) in the inference. Figure~\ref{fig:trajectory_comparison} shows an example of these trajectories recorded in the same unexperienced networks. We find that there is a turning point in the trajectory (at chunk 6 or 7 in Figure~\ref{fig:trajectory_comparison}) that IL- and RL-based models start to take an action that deviates from the optimal action by the expert policy. We speculate that it is because an unfamiliar observation is encountered when deciding on an action. More seriously, a bad chain reaction is observed for existing IL- and RL-based algorithms: a deviated action (from the expert one) is likely to transit the environment to a state with more unfamiliar observations, which in turn makes them take more deviated (sub-optimal) actions. As these algorithms only use past observations as input and utilize a typical neural network to fit a direct observations-to-action map for decision-making, they cannot break the vicious circles themselves, ultimately leading to significant performance degradation in the new environment. This may be a major cause of poor generalization for existing learning-based algorithms.

Therefore, a new ABR algorithm that can effectively control the deviation between its actions and the optimal trajectory is highly desired to achieve better generalization. Considering that an ABR task is essentially a causal sequential decision problem, 
we have an intuition that this new ABR algorithm should be able to capture the interrelated causality among observations, returns, and actions. For example, how past actions affect observations and returns, and how received observations and returns affect the decision of the next action. By taking the observations and returns as feedback signals, the ABR algorithm can analyze past actions and subsequently refine its next action for better feedback to catch up with the optimal action trajectory. As shown in Figure~\ref{fig:trajectory_comparison}, our proposed Karma equipped with causal sequence modeling can make timely action refinement after the occurrence of action deviation at chunk 7.

\vspace{-0.5cm}
\section{ABR algorithm via sequence modeling}
\label{sec:sequential_modeling}
To this aim, we propose Karma, which distinguishes itself from existing ABR algorithms via causal sequence modeling on past observations, returns, and actions to make decisions. This is achieved in two correlated ways:

\textit{1) Providing a multi-dimensional causal sequence as input.} We dynamically maintain a multi-dimensional causal sequence of past observations, returns, and actions to preserve the cross-time and inter-modality causality in state transitions. Karma uses this sequence as the input in both the training and inference stages. Specifically, the observation is a multidimensional signal, consisting of network throughput, buffer occupancy, and video information. The action signal indicates the bitrate to use for the next chunk.
The return signal, i.e., QoE-to-go, is defined as the maximum cumulative future QoE that can be attained to download all remaining chunks. Instead of a single-step instant QoE, such QoE-to-go allows Karma to comprehensively to better assess the current state from a long-term perspective. For example, the action with a high instant QoE may force the agent to take more conservative actions for subsequent chunks, which cannot be deemed a good one as the overall QoE for the entire video streaming session is degraded.

\textit{2) Applying a causal decision transformer for sequence modeling.} Empirical evidence suggests that a sequence modeling approach can model widely distributed behaviors, leading to a better generalization in the sequential-decision problems~\cite{hung2019optimizing,ramesh2021zero}. In Karma, a causal decision transformer is utilized to accurately characterize the state transition by modeling long-term dependencies among all three modalities, which inherently reside in sequential-decision ABR tasks. Compared to normal neural networks, this transformer architecture derives considerable advantages for effective sequence modeling from both the positional encoding and causal self-attention mechanism. The positional encoding provides unique position information for each input token, facilitating the transformer to discriminate between tokens at different positions. The causal self-attention mechanism empowers the transformer to concurrently attend to different tokens in the input sequence, capturing their causal relationship while ensuring that only the previous input information is utilized to predict the next chunk's action in line with the principle of causality. In this way, Karma learns to choose actions based on a comprehensive understanding of the causality among all modalities rather than just observations. 

\begin{figure}[t]
\centering
\setlength{\abovecaptionskip}{0.2cm} 
\includegraphics[scale = 0.54]{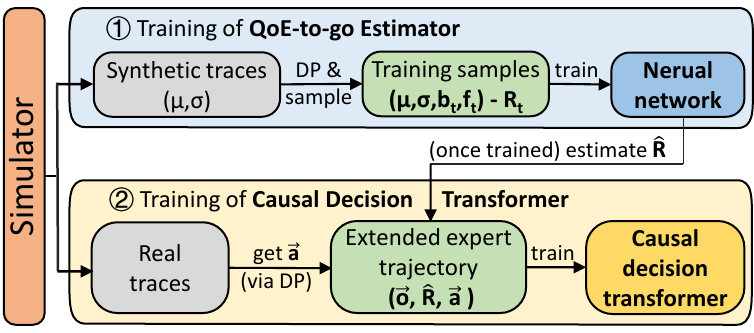}
\caption{The logical diagram of the training pipeline used by Karma.}
\label{fig:training_pipeline}
\vspace{-0.5cm}
\end{figure}

\section{Design of Karma}
\label{sec:design}

In this section, we describe the design and implementation of Karma. We first introduce the basic training algorithm of Karma. Then, we describe how to apply the trained Karma for adaptive bitrate selections during a video streaming session. Finally, we provide the implementation details of Karma.

\vspace{-0.2cm}
\subsection{Training Karma}

To train Karma, we must provide a set of extended expert trajectories as training samples, each containing a tuple of observations, corresponding QoE-to-go, and optimal action. To this aim, we introduce a simulator, as widely-used in~\cite{mao2017neural,pensievecode}, to simulate the video streaming environment faithfully, which largely accelerates the process of producing extended expert trajectories. As the accurate QoE-to-go is unavailable until the video streaming session ends, we first train a QoE-to-go estimator under synthetic network traces to generate the estimated QoE-to-go modality of extended expert trajectory recurrently based on current observations. Then we use the dynamic programming (DP) algorithm as an ABR method under real traces to generate the observation and corresponding action modalities of extended expert trajectory. Finally, a causal decision transformer is trained based on these extended expert trajectories. The logical diagram of the training pipeline is illustrated in Figure~\ref{fig:training_pipeline}.

\begin{figure}[t]
\centering
\setlength{\abovecaptionskip}{0.2cm} 
\includegraphics[scale = 0.7]{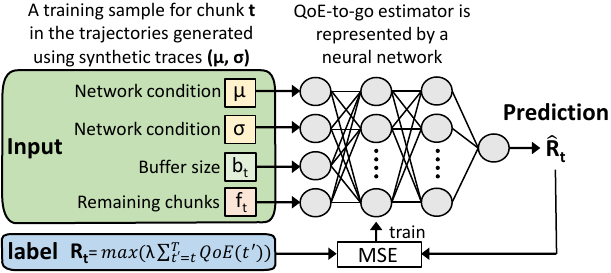}
\caption{The training algorithm for QoE-to-go estimator.}
\label{fig:rtg_estimator}
\vspace{-0.5cm}
\end{figure}

\noindent \textbf{Training QoE-to-go estimator: } As stated in the definition, the QoE-to-go at chunk $t$ (denoted as $R_t$) can be formulated as $R_t=max(\lambda\sum\limits_{t'=t}^T {QoE(t')})$. Once given the current buffer size $b_t$ and the chunk number $T$, $R_t$ is only determined by the network condition when downloading the remaining chunks. Based on the observation made by previous works~\cite{balakrishnan1997analyzing,zhang2001constancy} that end-to-end throughput is piece-wise stationary on a short-term time scale, we assume the network condition in the near future keeps the same as that currently. In this paper, we use the mean $\mu_{t}$ and the variance $\sigma_{t}$ of throughput values to characterize the network condition at time $t$ by following prior works~\cite{sun2016cs2p,akhtar2018oboe}. Hence, we suggest estimating the QoE-to-go based on the current observations, including the playback status and the network condition (e.g., the mean $\mu_{t}$ and the variance $\sigma_{t}$ of recent throughput measurements) as shown in Figure~\ref{fig:rtg_estimator}, and update the estimation chunk-by-chunk to adapt to possible new network conditions in the future.

\begin{figure}[t]
\centering
\setlength{\abovecaptionskip}{0.2cm} 
\includegraphics[scale = 0.52]{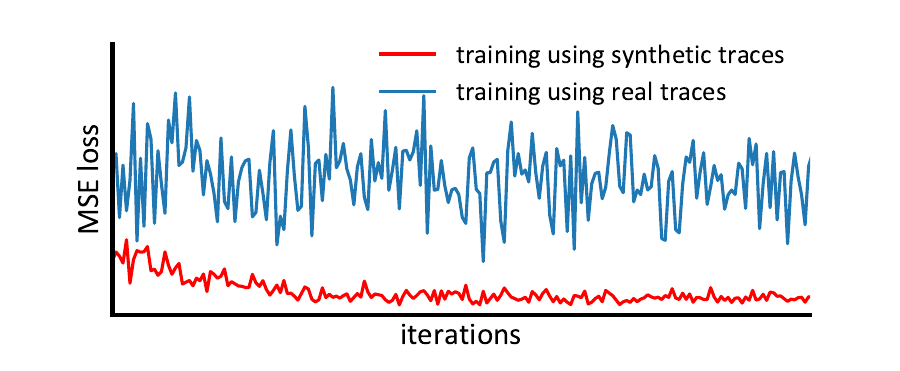}
\caption{Comparing the training loss of QoE-to-go estimator over time using synthetic traces with that using real traces.}
\label{fig:estimator_convergence}
\vspace{-0.45cm}
\end{figure}

To train the QoE-to-go estimator, we first produce a set of stationary synthetic traces, in which the throughput sample is generated following a Gaussian distribution with a predetermined mean $\mu$ and variance $\sigma$. This design ensures that the network condition remains relatively constant for all chunk downloads and guarantees the effectiveness of training samples for learning the mapping of current observations to QoE-to-go. It is worth noting that real traces with non-stationary bandwidth distributions do not meet the requirement of similar network conditions currently and in the future, so they are not suitable for the use of constructing training samples. Figure~\ref{fig:estimator_convergence} illustrates that the training of the QoE-to-go estimator can converge more effectively using synthetic traces than that using real traces.

Specifically, we generate training samples for the QoE-to-go estimator using these synthetic traces. Using DP as the ABR algorithm, the maximum cumulative QoE for all remaining chunks can be achieved, which is actually the truth value of QoE-to-go. As shown in Figure~\ref{fig:rtg_estimator}, we take a set of current observations, including the network condition $\mu$ and $\sigma$ (i.e., the settings for generating a synthetic trace), buffer size $b_t$, and the percentage of remaining chunks $f_t$ as the input, and use corresponding QoE-to-go $R_t$ as the label. Karma represents its QoE-to-go estimator as a two-layer fully connected network, and we train it by minimizing the mean squared error (MSE) loss between the output and the label.

\noindent \textbf{Training causal decision transformer: }
To ensure that Karma can gain experiences from the real environment, we use a broad set of network traces collected in the real world to train the causal decision transformer. For each observation ${\vec o}_t$, Karma uses network throughput measurements for the past $L$ chunks to compute $\mu_{t}$ and $\sigma_{t}$. Given the current observations of $\mu_{t}$, $\sigma_{t}$, $b_t$ and $f_t$, an estimated QoE-to-go ${\hat R}_t$ can be output by the well-trained QoE-to-go estimator. And an optimal action ${\vec a}_t$ is produced by using the dynamic programming algorithm. Once ${\vec a}_t$ is executed, Karma gets the next observation ${\vec o}_{t+1}$ and generates ${\hat R}_{t+1}$, which initiates a new round of action decision. 
In this way, the observation, estimated QoE-to-go, and optimal action together constitute an extended expert trajectory. For a video with a total of $T$ chunks, an extended expert trajectory can be expressed as a 3-modality sequence with $3T$ tokens in total:
\begin{equation}
  \tau  = \left( ({{{\vec o}_1}, {{\hat R}_1},{{\vec a}_1}),({{\vec o}_2},{{\hat R}_2}, {{\vec a}_2}),  \ldots ,({{\vec o}_T},{{\hat R}_T}},{{\vec a}_T}) \right),
\end{equation}
\noindent where $\vec{o_t}$, ${\hat R_t}$, and $\vec{a_t}$ respectively represent a set of observations, the estimated QoE-to-go, and the action for chunk $t$.
\begin{itemize}[leftmargin=*]
    \item Observation: At chunk $t$, the agent gets an observation vector
    \begin{math}
        {{\vec o}_t} = ({b_t},{c_t},{d_t},{{\vec e}_t} ,{f_t})
    \end{math} from the environment. $b_t$ is the buffer size; $c_t$ is the network throughput measurement; $d_t$ is the download time; ${{\vec e}_t}$ is a vector of all available sizes for the next video chunk; $f_t$ is the percentage of remaining chunks.
    \item Estimated QoE-to-go: ${{\hat R}_t}$, the estimation of maximum cumulative QoE for the remaining chunk downloads, is used in Karma to represent the return signal, which is generated from the trained QoE-to-go estimator.
    \item Action: ${{\vec a}_t}$ is a vector that indicates the probability distribution to select one from several available discrete bitrates for chunk $t$.
\end{itemize}

Then the causal decision transformer~\cite{chen2021decision}, a generative pre-trained transformer (GPT)~\cite{radford2018improving, ethayarajh2019contextual} model with causal self-attention masking, is introduced to execute sequence modeling and predict the bitrate for the next chunk. A training sample $(o, {\hat R}, a)$ is constructed by sampling a segment for $K$ consecutive tuple of observation, estimated QoE-to-go, and action from an extended expert trajectory. During each training epoch, a mini-batch of training samples is input to the transformer. Initially, we acquire token embeddings for timesteps and three modalities using a linear layer~\cite{ba2016layer}. Then, we perform positional encoding by embedding timestep to each token. In contrast to the conventional positional encoding scheme, where a single token corresponds to one timestep, Karma maps one timestep to three distinct tokens. These embeddings are then merged into interleaved tokens, which are further processed by multi-encoders and multi-decoders of the transformer to derive the hidden states. These encoders and decoders, equipped with causal self-attention masking, ensure that the current prediction in ABR tasks is not influenced by any information beyond the current timestep, i.e., an action $a_k$ can only be predicted using the $3k-1$ tokens before it in the sequence. Finally, the actions are predicted via another linear-layer decoder. The training loss comes from the cross-entropy of predicted actions and optimal actions, and the parameters of the Transformer are updated using the gradient descent algorithm. The training algorithm of the causal decision transformer is summarized in Algorithm~\ref{alg:basic_training_algorithm}. 

\begin{algorithm}[t]
  \SetKwData{Left}{left}\SetKwData{This}{this}\SetKwData{Up}{up}
  \SetKwFunction{Union}{Union}\SetKwFunction{FindCompress}{FindCompress}
  \SetKwInOut{Input}{input}\SetKwInOut{Output}{output}
  \tcp{\small $o,{\hat R},a,t$: observations, estimated QoE-to-go, actions, and timesteps}
  \tcp{\small$embed_t$:embedding for positional encoding}
  \tcp{\small$embed_o$,$embed_{\hat R}$,$embed_a$:linear embedding layers}
  \tcp{\small GPT: a transformer architecture with causal self-attention masking}
  \tcp{\small$pred_a$:linear action prediction layer}
  \BlankLine
  Create an initialized GPT\\
  \While{not reach max training episode}{
    Randomly pick a minibatch of $sample (o,{\hat R},a,t)$ from extended expert trajectories\\
    \For{sample ($o,{\hat R},a,t$) in minibatch}{\label{forins}
        \tcp{\small embeddings and positional encoding}
        \tcp{\small per-timestep with 3 tokens}
        $pos_{emb}$ = $embed_{t} ( t )$\\
        $o_{emb}$ = $embed_{o} ( o )$ + $pos_{emb}$\\
        ${\hat R}_{emb}$ = $embed_{\hat R} ( {\hat R} )$ + $pos_{emb}$\\
        $a_{emb}$ = $embed_a ( a )$ + $pos_{emb}$\\
        $input_{emb}$ = stack(${\hat R}_{emb}$, $o_{emb}$, $a_{emb}$)
        \BlankLine
        \tcp{\small get hidden states and predict next action}
        hidden\_states = GPT($input_{emb}$)\\
        $a_{hidden}$ = unstack(hidden\_states).actions\\
        $a_{preds}$ = $pred_a$($a_{hidden}$)\\
        \BlankLine

        \tcp{\small Cross-Entropy for discrete actions}
        loss = CrossEntrophy($a_{preds}$, $a$)\\
        optimizer.zero\_grad() \\
        loss.backward() \\
        optimizer.step()
    }
  }
  \caption{}
  \label{alg:basic_training_algorithm}
\end{algorithm}\DecMargin{1em}

\subsection{Inference}
\begin{figure}[t]
\centering
\setlength{\abovecaptionskip}{0.2cm} 
\includegraphics[scale = 0.80]{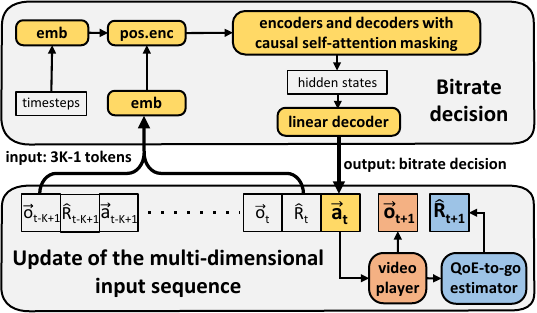}
\caption{The logical diagram of the inference pipeline used by Karma.}
\label{fig:online_pipeline}
\vspace{-0.5cm}
\end{figure}

\noindent \textbf{Bitrate decision:} To select a bitrate for video chunk $t$, Karma first utilizes a multi-dimensional sequence of observations, estimated QoE-to-go, and actions for the past $K$ chunks as input, which totally possess $3K-1$ tokens (not including $a_t$ token which is to be predicted). These tokens are then combined with their timesteps to execute positional encoding. Finally, Karma passes the positional-encoded tokens into the trained causal decision transformer to predict the bitrate ${\vec a}_t$ of the video chunk $t$. Figure~\ref{fig:online_pipeline} illustrates the logical diagram of the inference pipeline used by Karma.

\noindent \textbf{The update of input sequence:} Once the video player executes ${\vec a}_t$, the environment transits to a new state with the observation ${\vec o}_{t+1}$ and an estimated QoE-to-go ${\hat R}_{t+1}$. Similar to training, ${\hat R}_{t+1}$ is generated via the trained QoE-to-go estimator using $b_{t+1}$ and $f_{t+1}$ from ${\vec o}_{t+1}$, as well as mean $\mu_{t+1}$ and variance $\sigma_{t+1}$ calculated using the past $L$ network throughput measurements. Obviously, the estimated QoE-to-go is updated chunk-by-chunk recurrently, which can improve estimation accuracy when the network condition or playback status changes. Subsequently, ${\vec a}_t$, ${\vec o}_{t+1}$, and ${\hat R}_{t+1}$ are all added into the multi-dimensional sequence, and the oldest ${\vec o}_{t-K+1}$, ${\hat R}_{t-K+1}$, and ${\vec a}_{t-K+1}$ are removed from the sequence. This update process continues until the end of the video. 

\subsection{Implementation}
In the implementation of Karma, we use the past $L=4$ chunks to characterize the current network condition for the QoE-to-go estimator and use a multi-dimensional sequence for the past $K=4$ chunks as the input of the transformer. The impact of $L$ and $K$ on Karma's performance will be discussed in §\ref{ssec:ablation_peformance}. Similar to Oboe\cite{akhtar2018oboe,oboereproduce}, we generate synthetic traces with mean $\mu$ ranging from 0.5Mbps to 6Mbps in increments of 0.1Mbps and variance $\sigma$ ranging from 0 to 3 in increments of 0.1, to produce training samples for QoE-to-go estimator. The scale factor $\lambda$ is empirically set to 0.01. The transformer uses a neural network structure of 3 hidden layers, one attention head, and a 128-dimensional embedding operation. 
During training, we use an initial learning rate of 0.001 and then dynamically adjust it using the cosine decay manner. We update the parameters using the AdamW optimizer~\cite{kingma2014adam}. Relu is used as the activation function, and the dropout is set to 0.1. The mini-batch size is set to 128. We use PyTorch to implement the QoE-to-go estimator and the Transformer architectures. After Karma generates an ABR algorithm, it is necessary to apply it to real-world video streaming sessions. For this purpose, we deploy Karma on a separate ABR server for real-world applications, which is implemented using the Python $BaseHTTPServer$.

\section{Evaluation}
\label{sec:evaluation}

\begin{figure*}[ht]
    \vspace{-0.5cm}
    \subfigcapskip=-4pt
    \centering
    \subfigure[The CDF distributions of general QoE]{\includegraphics[scale = 0.36]{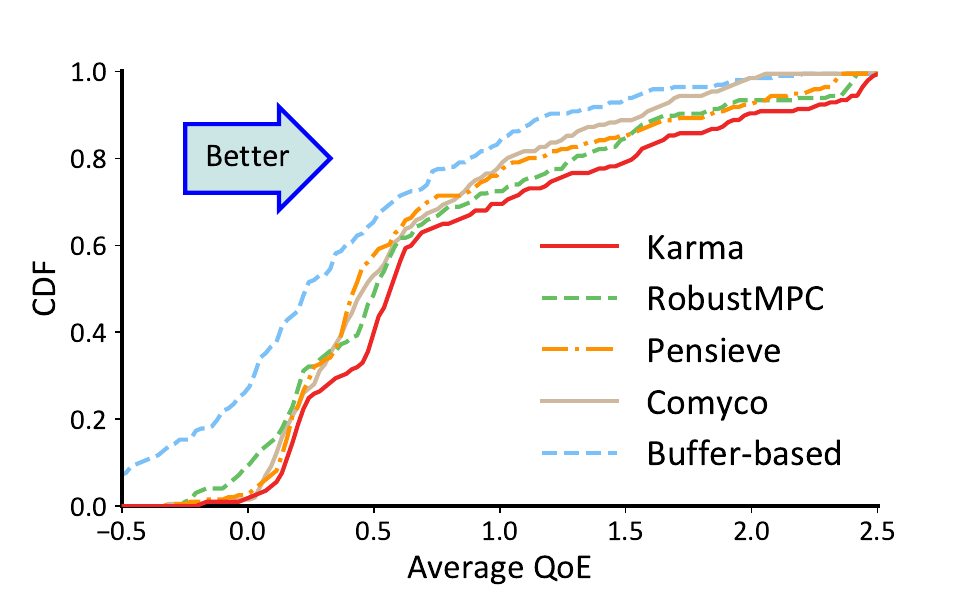}}\label{sfig:eval_FCC_QoE}
    \hspace{4mm}
    \subfigure[The average value of general QoE metric and its individual components]{\includegraphics[scale = 0.45]{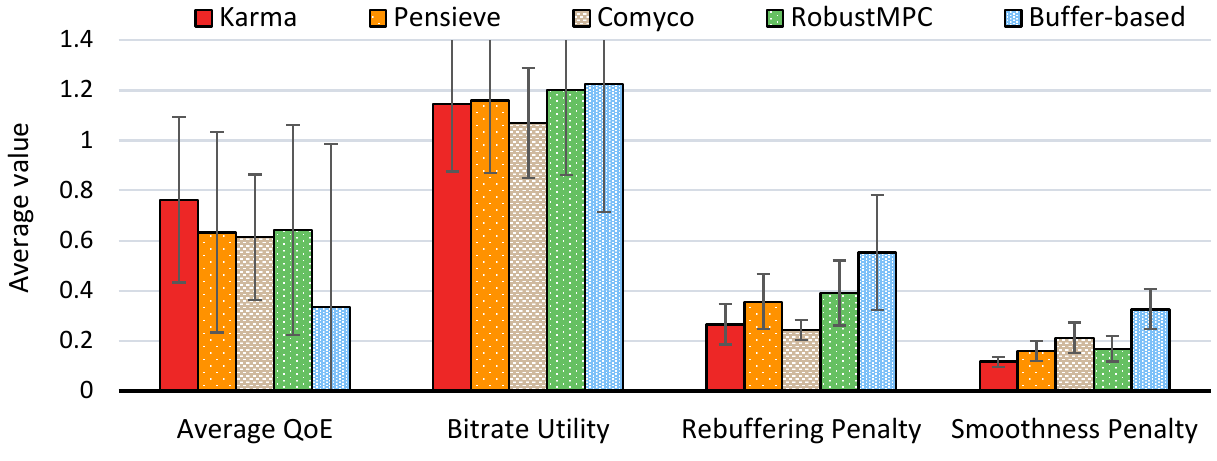}}\label{sfig:eval_FCC_bar}
    \vspace{-5mm}
    \caption{Comparing Karma with existing ABR algorithms on FCC broadband networks.
    The mean and variance of throughput in the FCC dataset are 1.30 and 0.99 respectively.
    }
    \label{fig:eval_FCC}
    \vspace{-0.4cm}
\end{figure*}

\begin{figure*}[ht]
    \vspace{-0.2cm}
    \subfigcapskip=-4pt
    \centering
    \subfigure[The CDF distributions of general QoE]{\label{sfig:eval_norway_QoE}{\includegraphics[scale = 0.36]{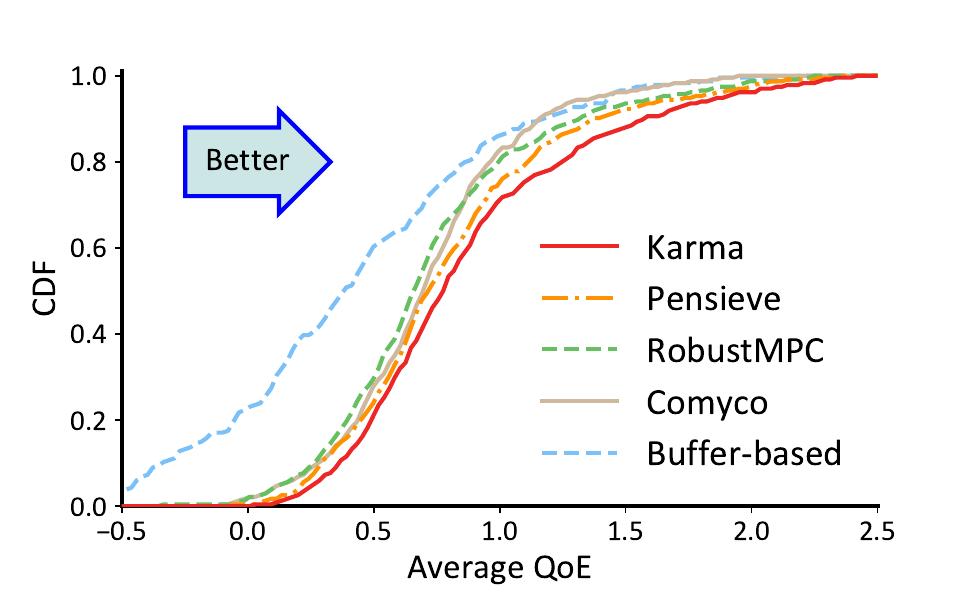}}}  
    \hspace{4mm}
    \subfigure[The average values of general QoE and its individual components]{\includegraphics[scale = 0.45]{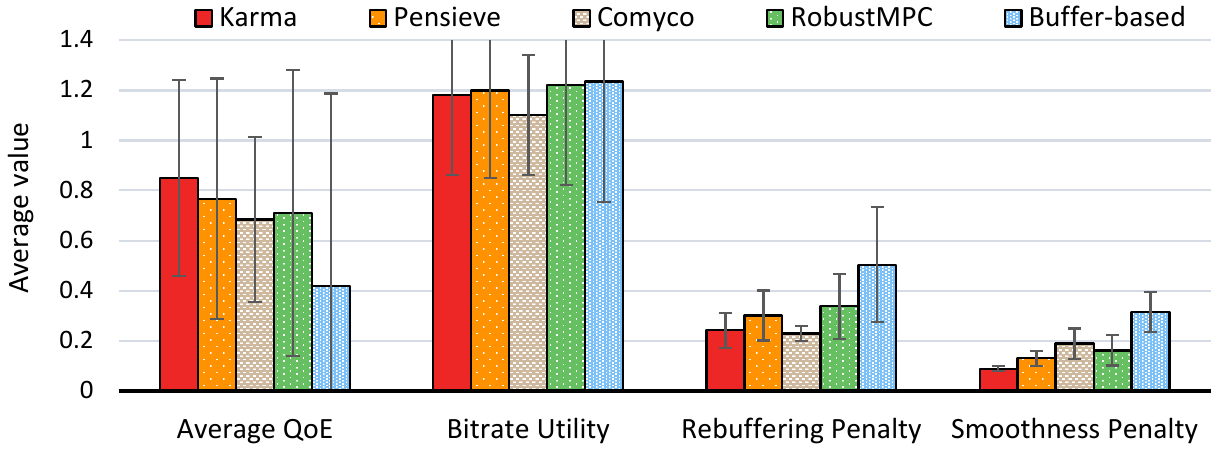}}\label{sfig:eval_norway_bar}
    \vspace{-5mm}
    \caption{Comparing Karma with existing ABR algorithms on Norway 3G/HSDPA networks. 
    The mean and variance of throughput in the Norway dataset are 1.56 and 0.97 respectively.
    }
    \label{fig:eval_norway}
    \vspace{-0.4cm}
\end{figure*}

\subsection{Methodology}

\noindent \textbf{Network traces:} 
To evaluate Karma on realistic network conditions, we use real network traces from several popular public datasets, including FCC's broadband dataset~\cite{coase2013federal}, and 3G/HSDPA mobile dataset~\cite{riiser2013commute} collected in Norway. We use 70$\%$ of these traces
as a training set and the remaining 30$\%$ as a test set. Another Oboe dataset~\cite{akhtar2018oboe}, which collects traces from wired, WiFi, and cellular network connections, is introduced only for validation. 
Each trace is meticulously filtered to satisfy the requirement of average throughput below 6Mbps and minimum throughput above 0.2Mbps. 
Moreover, all traces were formatted to be compatible with the Mahimahi network simulation tool. 

\noindent \textbf{QoE metric:} 
Due to the preferences of different users ~\cite{ketyko2010qoe,mok2011inferring,mok2011measuring,piamrat2009quality}, we use a general form of linear QoE metric for video chunk $t$, which is defined as
\begin{equation}
  QoE(t) = q({r_t}) - \eta {T_t} - \gamma \left| {q({r_{t}}) - q({r_{t-1}})} \right|
\label{eq:QoE}
\end{equation}
where $r_t$ represents the selected bitrate for chunk $t$, $q({r_t})$ maps the bitrate to the quality perceived by a user, and $T_t$ represents the rebuffering time. $\eta$ and $\gamma$ are penalty factors for rebuffering and quality smoothness loss. In accordance with recent researches~\cite{mao2017neural,yuan2022prior,kan2022improving}, we choose to set $q({r_t}) = 0.001r_t$ ($r_t$ in kbps), $\eta = 4.3$ and $\gamma = 1.0$.

\BlankLine
\noindent \textbf{ABR Baselines: } 
\begin{itemize}[leftmargin=*]
    \item Pensieve~\cite{mao2017neural}: firstly introduces RL-based techniques into adaptive video streaming applications. Its return modality is only used in the training stage. In this paper, we use the pre-trained Pensieve model provided by the authors.
    \item Comyco~\cite{huang2019comyco}: a typical IL-based ABR algorithm. It learns the control policy for bitrate adaptation by imitating the observation-to-action behavior of an expert. In this paper, we retrain Comyco using the QoE metrics defined in Equation~\eqref{eq:QoE}.
    \item RobustMPC~\cite{yin2015control}: MPC solves a QoE maximization problem over a horizon of several future chunks based on buffer occupancy and throughput prediction to select bitrates. Furthermore, RobustMPC introduces a normalized error between the predicted and actual throughput to make the algorithm more robust.
    \item Buffer-based (BB)~\cite{huang2014buffer,spiteri2020bola}:  dynamically selects the next bitrate according to the current buffer occupancy.
\end{itemize}

\noindent \textbf{Experimental setup:} The experimental setup is consistent with Pensieve~\cite{mao2017neural}. Here we give a brief description. The ``Envivio-
Dash3'' video was used for evaluation, which was divided into 48 chunks. Each chunk represents approximately 4 seconds of video playback time. The video provides six discrete bitrates as $\{$300, 750, 1200, 1850, 2850, 4300$\}$ kbps, which pertain to video resolutions in $\{$240, 360, 480, 720, 1080, 1440$\}$p. A playback buffer capacity of 60 seconds was configured for the player.
Please refer to Pensieve~\cite{mao2017neural} for details.

\begin{figure*}[ht]
    \vspace{-0.5cm}
    \subfigcapskip=-4pt
    \centering
    \subfigure[The CDF distributions of general QoE]{\includegraphics[scale = 0.36]{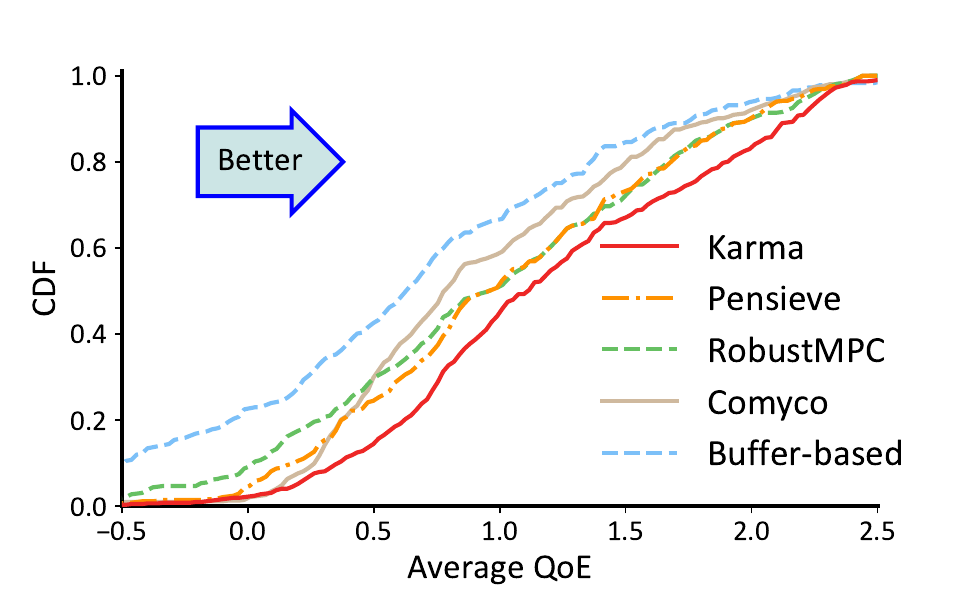}}\label{sfig:eval_oboe_QoE}
    \hspace{4mm}
    \subfigure[The average values of general QoE and its individual components]{\includegraphics[scale = 0.45]{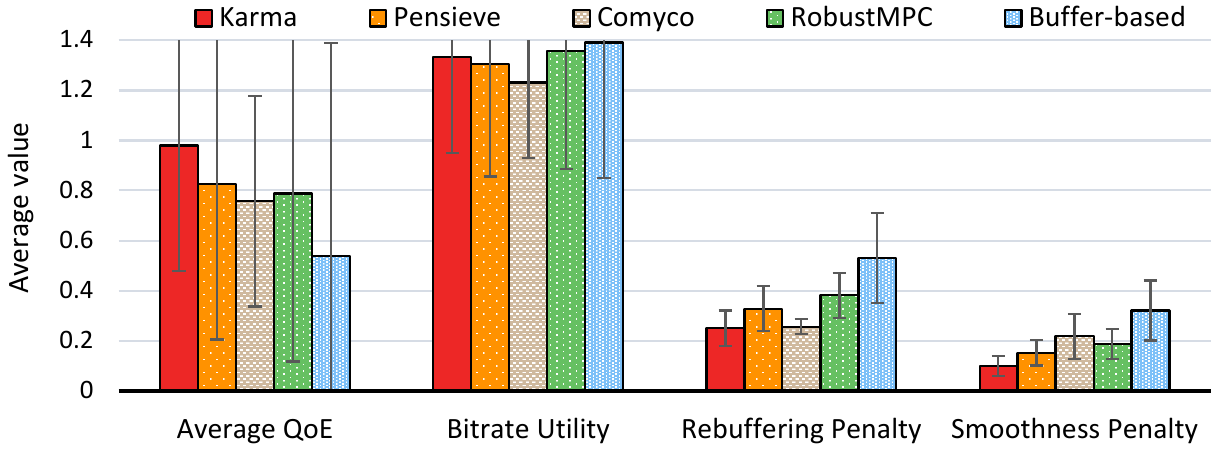}}\label{sfig:eval_bar_oboe}
    \vspace{-5mm}
    \caption{Comparing Karma with existing ABR algorithms on unexperienced Oboe network traces. 
    The mean and variance of throughput in the Oboe dataset are 1.79 and 1.36 respectively.
    }
    \label{fig:unexperienced_networks}
    \vspace{-0.15cm}
\end{figure*}

\begin{figure}[t]
\centering
\includegraphics[scale = 0.40]{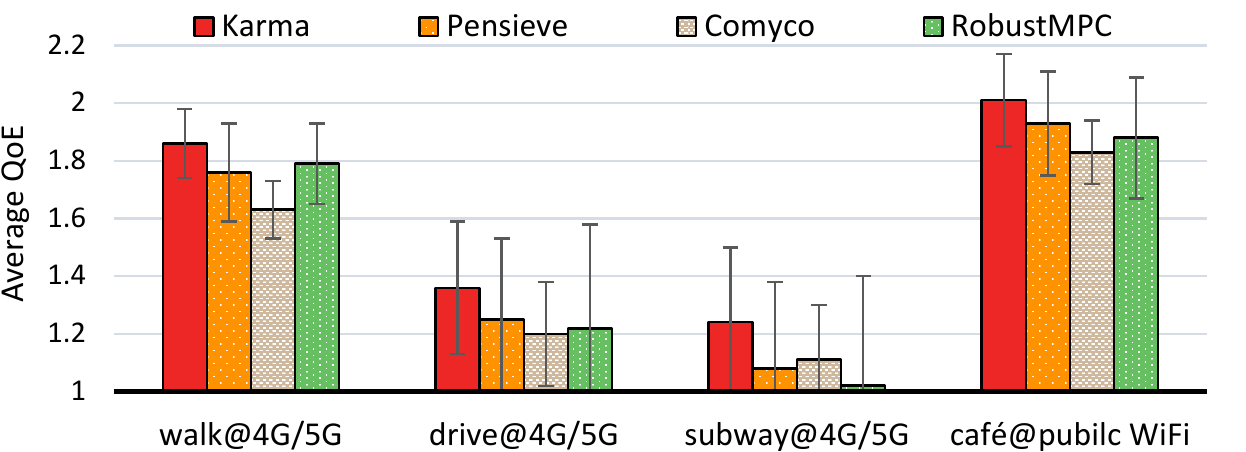}
\vspace{-7mm}
\caption{Comparing Karma with existing ABR algorithms in the wild. Results are collected on the 4G/5G cellular and public WiFi networks in the scenes of walking, driving, subway, and stationary caf\'e.}
\label{fig:eval_real_exp}
\vspace{-0.5cm}
\end{figure}

\vspace{-0.2cm}
\subsection{Karma versus Existing ABR Algorithms}
\label{ssec:overall_performance}
To evaluate Karma, we compare it with different types of state-of-the-art ABR algorithms. Figure~\ref{fig:eval_FCC} and Figure~\ref{fig:eval_norway} depict the results in the form of cumulative distribution function (CDF) distributions and average values of general QoE on the FCC broadband dataset and the Norway 3G/HSDPA dataset respectively. The average values of individual components in the general QoE definition (Equation~\eqref{eq:QoE}) are also provided, including bitrate utility, rebuffering penalty, and smoothness penalty.

In general, Karma has shown to be a superior ABR algorithm compared to the existing solutions, with an average QoE improvement ranging from 10.8$\%$ (Pensieve on Norway networks) to 18.7$\%$ (RobustMPC on FCC networks). As evidenced by the CDF distribution, Karma is able to adapt to various network conditions, which is a strong demonstration of robustness. Karma does not lead in every underlying metric when compared to other ABR algorithms. For example, Pensieve and RobustMPC usually exhibit a better average bitrate utility. However, Karma exhibits better control of rebuffering events and frequent bitrate switching, which helps it stand out from all schemes.

The closest competing schemes are Pensieve and RobustMPC. Comyco employs a policy that tends to choose a lower bitrate and perform frequent bitrate switches to ensure smooth video playback. This conservative policy only helps to reduce rebuffering events but does not necessarily lead to a satisfactory overall QoE. BB falls behind other schemes significantly because it makes bitrate decisions only based on a fixed rule of current buffer occupancy, making it struggle to adapt to different network conditions.

\begin{figure*}[h]
    \centering
    \begin{minipage}{0.29\linewidth}
    \vspace{-5mm}
    \includegraphics[width=\linewidth]{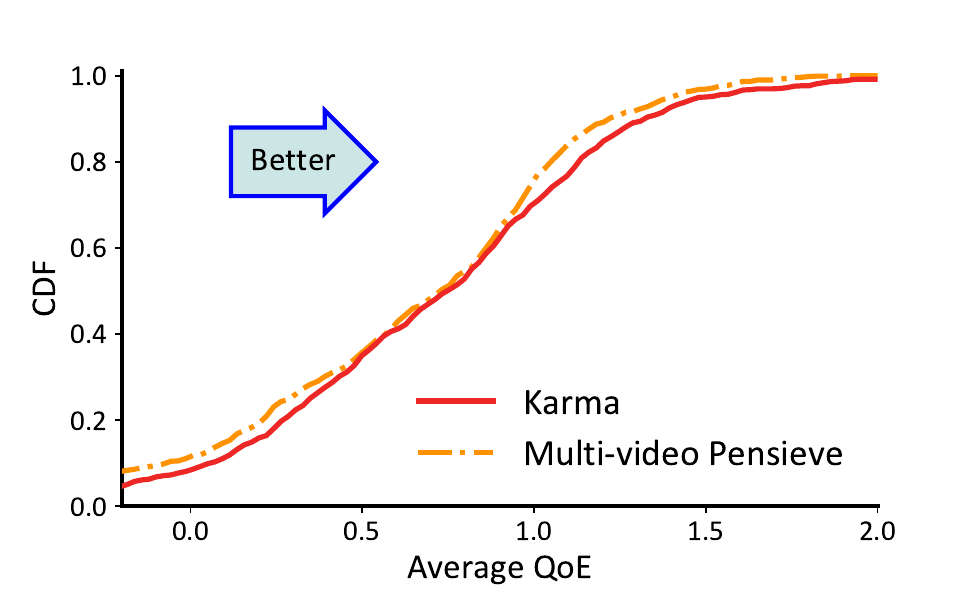}
    \vspace{-7mm}
    \caption{Comparing Karma with multi-video Pensieve across multiple video properties.}
    \label{fig:eval_multi_video}
    \end{minipage}
    \hspace{0.7cm}
    \begin{minipage}{0.29\linewidth}
    \vspace{-5mm}
    \includegraphics[width=\linewidth]{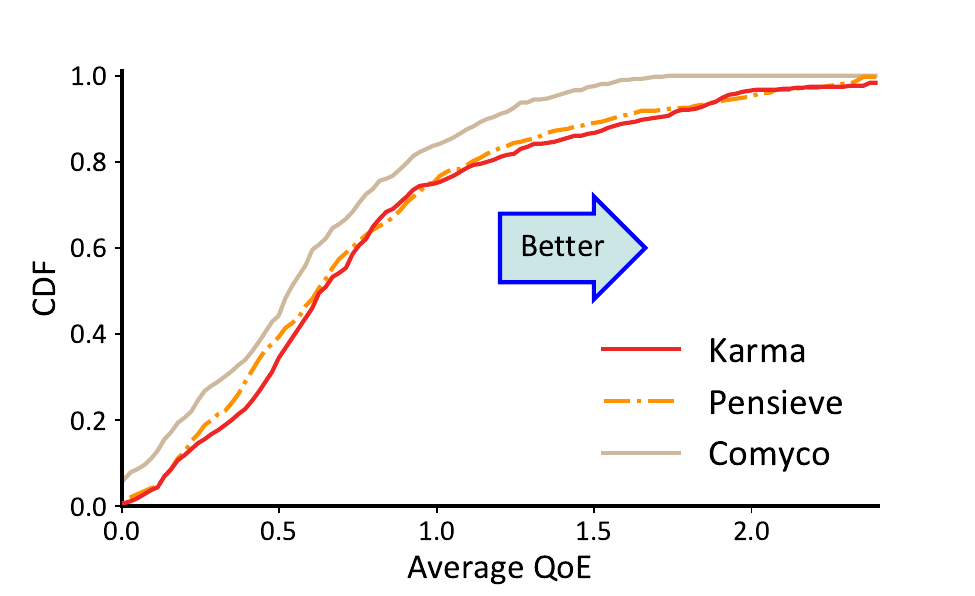}
    \vspace{-7mm}
    \caption{Comparing Karma with Comyco when trained using Pensieve as the expert policy.}
    \label{fig:sub-optimal_expert_policy}
    \end{minipage}
    \hspace{0.7cm}
    \begin{minipage}{0.29\linewidth}
    \vspace{-5mm}
    \includegraphics[width=\linewidth]{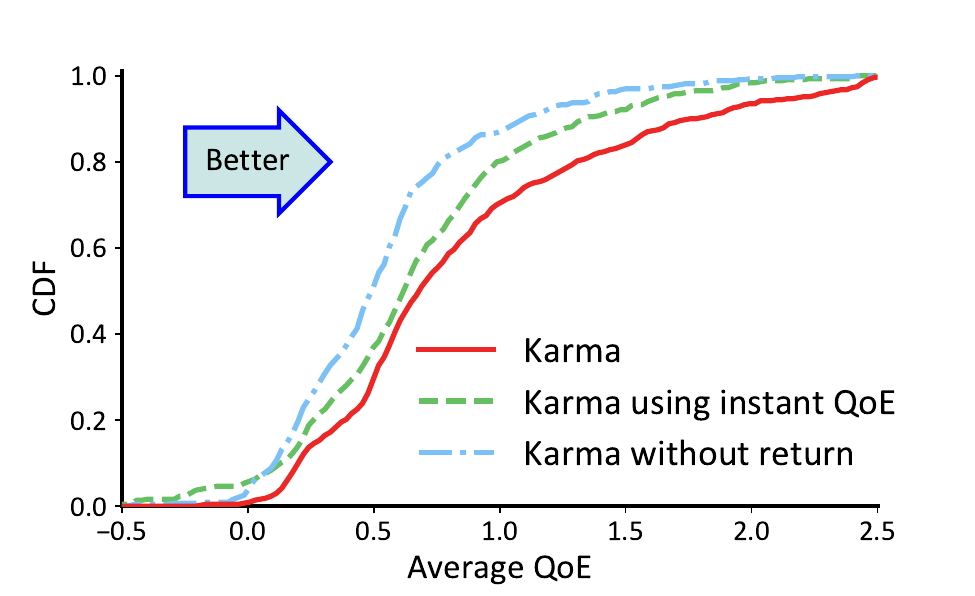}
    \vspace{-7mm}
    \caption{Comparing Karma with its variants using different forms of the return signal.}
    \label{fig:form_of_return_signal}
    \end{minipage}
    \vspace{-0.3cm}
\end{figure*}

\vspace{-0.2cm}
\subsection{Generalization}
\label{ssec:generalization_peformance}
To evaluate the generalization of Karma, we first conduct several experiments across a wide range of network conditions in both simulation and the real world and across multiple video properties. Second, we retrain Karma with an expert policy that is sub-optimal but more easily available and evaluate its effectiveness. 

\textbf{Unexperienced network traces}: We compare Karma with existing ABR algorithms under the network traces in the Oboe dataset, which has never been experienced for all learning-based ABR algorithms in the training stage. The Oboe dataset has a different distribution of network conditions, bringing more challenges for these ABR algorithms to retain their performance. As shown in Figure~\ref{fig:unexperienced_networks}, we find that Karma achieves the best generalization under the unexperienced network conditions, with a stable improvement of 19.2$\%$ over the second-best Pensieve.

\textbf{Real-world field tests}: We conduct several experiments in the wild to evaluate Karma and several state-of-the-art ABR algorithms (Pensieve, Comyco, and RobustMPC) on 4G/5G cellular and public WiFi networks in the scenes of walking, driving, subway, and stationary caf\'e. The evaluation is carried out on a client running on Ubuntu 18.04 who performs actual video requests and downloads from a video server hosted on a node of a 3rd-party Internet cloud. In these experiments, we load the test video ten times in random order using each ABR algorithm. As depicted in Figure~\ref{fig:eval_real_exp}, Karma always achieves the best performance in the form of average QoE on different networks and in different scenes. Over relatively stationary network connections, e.g., walking@4G/5G cellular network and caf\'e@public WiFi network, Karma has a slight improvement of about 4$\%$ compared with the second-place scheme (RobustMPC in walking@4G/5G and Pensieve in caf\'e@public WiFi). While over more dynamic network connections, e.g., driving@4G/5G cellular network and subway@4G/5G cellular network, Karma significantly outperforms other ABR algorithms by 7.1$\%$-11.7$\%$. The network dynamics are mainly caused by high mobility and frequent handovers~\cite{lin2022bandwidth,rikic2021cellular}. Interestingly, we also find that existing ABR algorithms show inconsistent performances in different scenes. For example, Pensieve, RobustMPC, and Comyco respectively rank second in the scene of driving@4G/5G (and caf\'e@public WiFi), walking@4G/5G, and subway@4G/5G. It reveals that Karma can generalize well to different networks while existing ABR algorithms cannot.

\textbf{Multiple videos}: We also evaluate Karma on other videos different from the one in training, which are very common in real life. For each test, a video was generated synthetically with different properties including bitrate ladders, number of chunks, and chunk size. Specifically, the number of available bitrates was randomly selected from $[4,8]$, and the number of chunks was randomly selected from $[30,60]$. Other properties, including the chunk size, were determined using the method presented in Pensieve~\cite{mao2017neural}. 
For the video not with 6 bitrate ladders, Kamra just maps its predicted action down to an available bitrate without retraining. The Multi-video Pensieve, a retrained ABR model on the above-mentioned synthetic videos, is used for comparison. Figure~\ref{fig:eval_multi_video} shows that Karma outperforms Multi-video Pensieve with an improvement of 5.5$\%$. This result demonstrates that Karma can generalize well across different video properties.

\textbf{Training using sub-optimal expert trajectories}: Considering that the absolute optimal expert trajectories are sometimes unavailable (needing known network traces) or costly (high complexity in dynamic programming) to acquire, we seek to validate the effectiveness of Karma 
when only a feasible sub-optimal expert policy can be provided. Specifically, we use Pensieve to generate a series of sub-optimal expert trajectories (compared to that generated via dynamic programming) for training Karma as well as IL-based Comyco. Figure~\ref{fig:sub-optimal_expert_policy} shows that Karma outperforms Comyco significantly with the improvements on average QoE of 28.8$\%$. It implies that Karma can generalize effectively even if the expert policy is sub-optimal. Notably, Karma even exceeds the expert trajectories themselves (i.e., Pensieve's performance) with a slight improvement of 2.7$\%$. An explanation is that by generating reasonable QoE-to-go tokens using the estimator, Karma can execute timely action refinement to get close to or even exceed the desired target. 
Thus, even if trained with the guidance of a sub-optimal policy like Pensieve, this inherent mechanism helps Karma partly alleviate the deviation from optimal action in the inference, which existing ABR algorithms cannot achieve.

\vspace{-0.3cm}
\subsection{Karma Deep Dive}
\label{ssec:ablation_peformance}
In this section, we set up a series of experiments to gain a thorough understanding of Karma. Specifically, we investigate the optimal input sequence length (i.e., $K$) for the transformer, the optimal window size of past chunks (i.e., $L$) used to generate network condition features for the QoE-to-go estimator and the most effective form of the return signal in Karma. 

\textbf{The input sequence length}: We have tested Karma's performances using different $K$. As shown in Table~\ref{tab:input_length}, the best performance is achieved when $K$ is 4. However, Karma does not deteriorate significantly when $K$ varies within a reasonable range. For instance, the performance drop is limited to 5$\%$ when $K$ is either 3 or 5. But if the input sequence is too short (e.g., $K=1$), Karma fails to capture the effective causality, leading to a severe performance collapse. Meanwhile, a higher value of $K$ (e.g., $K=8$) also does not necessarily guarantee a performance improvement, implying that excessive, unnecessary historical information may interfere with Karma's decision. 

\par\textbf{The window size for QoE-to-go estimator}: We also study the effect of using varying $L$ for the QoE-to-go estimator on the performance of Karma. Results are listed in Table~\ref{tab:past_chunk_number_for_rtg_estimator}. The best performance is attained for Karma when $L$ is set to 4. Because the network condition usually fluctuates over time, it's insufficient only to use a single throughput sample (e.g., $L=1$) to represent the current network condition. On the contrary, a throughput series that contains information from a long time ago (e.g., $L=8$) also fails to accurately represent the current network condition, which even results in a worse performance of Karma.

\textbf{The form of return signal}: 
To verify how QoE-to-go benefits Karma, we develop two variants of Karma for comparison: one using instant QoE as the return signal and the other learning from the expert trajectories without any return signal. The results are illustrated in Figure~\ref{fig:form_of_return_signal}.
We find that the average QoE for Karma is 25.4$\%$ higher than that using instant QoE. We conjecture that the instant QoE can hardly establish proper causality among past modalities for sequence modeling, finally preventing Karma from learning appropriate policy in a causal sequence modeling way.
Similarly, the scheme that ignores the return signal seems like a kind of simple behavior cloning and falls behind Karma with a vast gap of 42.7$\%$.

\textbf{Karma Overhead}: 
We train and test Karma on a 12-core, AMD R5-4600H 3.00Hz CPU. Because of the direct expert policy, Karma can be proficiently trained within 2 hours (similar to Comyco and faster than Pensieve). The model size is 5.3 MB, and the decision-making time is in milliseconds. In short,
we believe that Karma does not incur significant computational overhead and is highly feasible for practical implementation.

\vspace{-0.2cm}
\section{Related works}

\noindent \textbf{ABR algorithms:} Most early ABR algorithms develop fixed rules based on environmental observations. For example, buffer-based (BB)~\cite{huang2014buffer,spiteri2020bola} and rate-based (RB)~\cite{sun2016cs2p,jiang2012improving} choose bitrate based on buffer occupancy and estimated network throughput, respectively. MPC\cite{yin2015control}, 
as the state-of-the-art rule-based approach, uses both buffer occupancy information and estimated throughput to select bitrate by solving a QoE maximization problem over a horizon of several future chunks. Emerging learning-based ABR algorithms can be classified into IL-based algorithms~\cite{huang2019comyco} and RL-based algorithms~\cite{mao2017neural,huang2018tiyuntsong,gadaleta2017d}. They benefit from the excellent fitting ability of neural networks and can learn a better ABR policy from expert trajectories or through exploration. However, they typically rely on a direct observations-to-action map for decision-making and tend to generalize poorly in a new environment with unfamiliar observations.

\begin{figure}[t]
\vspace{-0.3cm}
\centering
\begin{minipage}[t]{\textwidth}
\hspace{-0.6cm}
 \begin{minipage}[t]{0.25\textwidth}
  \centering
        \makeatletter\def\@captype{table}\makeatother\caption{Different settings of $K$ on Karma's performance.}
\begin{tabular}{cc} \hline 
\toprule 
$K$ & Average QoE  \\ 
\midrule 
1 & 0.564 $\pm$ 0.095  \\
3 & 0.762 $\pm$ 0.031  \\
\textbf{4} & \textbf{0.793 $\pm$ 0.021}  \\
5 & 0.774 $\pm$ 0.028  \\
8 & 0.690 $\pm$ 0.036  \\
\bottomrule 
\end{tabular}
\label{tab:input_length}
  \end{minipage}
  \hspace{0.1cm}
  \begin{minipage}[t]{0.25\textwidth}
   \centering
        \makeatletter\def\@captype{table}\makeatother\caption{Different settings of $L$ on Karma's performance.}
         \begin{tabular}{cc} \hline 
\toprule 
$L$ & Average QoE  \\ 
\midrule 
1 & 0.687 $\pm$ 0.144  \\
3 & 0.774 $\pm$ 0.023  \\
\textbf{4} & \textbf{0.793 $\pm$ 0.021}  \\
5 & 0.744 $\pm$ 0.042  \\
8 & 0.511 $\pm$ 0.109  \\
\bottomrule 
\end{tabular}
\label{tab:past_chunk_number_for_rtg_estimator}
   \end{minipage}
\end{minipage}
\vspace{-0.4cm}
\end{figure}

\noindent \textbf{Sequence modeling in decision problem:}
Recently, a new technology paradigm that applies sequence modeling using the transformer~\cite{vaswani2017attention} to solve decision-making problems appears. It stands out from traditional learning approaches due to its remarkable capability of modeling long sequences. 
For instance, in some essential RL decision scenarios, such as Gym~\cite{brockman2016openai} and Atari~\cite{bellemare2013arcade}, Decision Transformer (DT)~\cite{chen2021decision} meets or even exceeds the best temporal difference learning-based traditional RL. Similarly, Trajectory Transformer (TT)~\cite{janner2021offline} models the distribution over trajectories as a planning algorithm. However, as far as we know, there is no research that directly applies causal sequence modeling to the ABR task.

\section{Conclusion}
This paper proposed Karma, an ABR algorithm that applied causal sequence modeling for ABR optimization. Karma first maintained a multi-dimensional causal sequence of past observations, QoE-to-go, and actions as input and then applied a causal decision transformer for causal sequence modeling and final bitrate decision. Experimentally, Karma outperformed existing fixed rule-based and learning-based ABR algorithms, with an average QoE improvement of 10.8$\%$-18.7$\%$. Karma also proved its ability to generalize well across various networks and multiple video properties in both simulations and real-world field tests.

\begin{acks}
This work was supported by the National Natural Science Foundation of China (62101241, 62231002) and the Jiangsu Provincial Double-Innovation Doctor Program (JSSCBS20210001). {\it (Corresponding Author: Hao Chen.)}
\end{acks}

\newpage
\bibliographystyle{ACM-Reference-Format}
\balance
\bibliography{reference}

\end{sloppypar}
\end{document}